**TITLE:** Alternative to the statistical mass confusion of testing for “no effect”

**RUNNING TITLE:** P-values answer the wrong question

**ETOC STATEMENT:** Progress in cell biology is hindered by the near-universal practice of basing conclusions on statistical tests of the no-effect hypothesis. A modern, quantitatively rigorous cell biology requires that distinctions between strong and weak effects are built into how we analyze our data.

**AUTHOR:** Josh L. Morgan

**AFFILIATION:** Washington University in St. Louis, Department of Ophthalmology and Visual Sciences, Neuroscience, Biology and Biomedical Science.

**CONTACT:** jlmorgan@wustl.edu

## ABSTRACT

It should not be controversial to argue that the proximate goal of measuring something is to figure out how big or small or fast or slow it is. Estimates of effect size can be used to build models of how cells work and to test quantitative predictions. Unfortunately, in cell biology, quantification is nearly synonymous with null-hypothesis significance testing. The hypothesis being tested is universally assumed to be the hypothesis that there was no effect. Framing every experiment as an attempt to reject the no-effect hypothesis is convenient but doesn't teach us about cells.

In this manuscript, I walk through some of the common critiques of significance testing and how these critiques relate to experimental cell biology. I argue that careful consideration of effect size should be returned to its central position in the planning and discussion of cell biological research. To facilitate this shift in focus, I recommend replacing p-values with confidence intervals as cell biology's default statistical analysis.

## MAIN TEXT

Biologists collect measurements that can be used to inspire, refine, or distinguish between models of the world. We can use descriptive statistics to extract simplifying parameters from these measurements. We can then use statistical models to understand the precision of our measurements. Rigorous reporting of measurements should, therefore, include two classes of information: *magnitude* (mean, median, correlation coefficient, etc.) and *precision* (standard error, confidence interval, compatibility interval).

Statistical analysis can also provide a third class of information, quantification of the difference between the observed result and a statistical model of a hypothesis. To perform this quantification, a researcher chooses a model that approximates both the hypothetical behavior of a biological measure and the researcher's sampling of that measure. This statistical model can be a readily parameterized set of normal distributions or a complex computer simulation. By comparing the model's predictions to our observed results, we can evaluate the extent to which our observations are compatible with our assumptions and hypotheses.

To quantify the relationship between a modeled hypothesis and an observed result, most cell biologists use a p-value. $P = 0.01$ can be interpreted as follows: 1) If the assumptions of the statistical model exactly fit the experiment and 2) if the test (null) hypothesis exactly describes the underlying biology, then 3) a result at least as different from the test hypothesis as the observed result would be expected in 1% of an infinite number of similar experiments (Wasserstein and Lazar, 2016). A low p-value (<0.05 by convention) is usually interpreted as evidence against the null hypothesis. While many of the problems with this practice are discussed below, it is worth emphasizing immediately that $p = 0.01$ absolutely DOES NOT mean there is a 1% chance that the null hypothesis is an accurate description of the world.

In a narrow set of experiment types (Szucs and Ioannidis, 2017), null hypothesis significance testing (NHST) can be useful (not ideal) in probing data. For instance, if a single preliminary experiment reveals 3,000 genes that increase their expression level after injury, p-values might help you decide which genes to follow up on first. However, a p-value should never be interpreted as a measurement of a biological property. The p-value is, at best, a single, non-intuitive fragment of a decision-making process about specific sets of experiments.

Problems with significance testing have been pointed out since it was popularized by R. A. Fisher a hundred years ago (Rozeboom, 1960; Goodman, 1993; Meehl, 1967; Berkson, 1938; Hogben, 1957; Bakan, 1966; Sterling, 1959; Calin-Jageman, 2022; Morrison, 1970). Misinterpretations of p-values range from relatively subtle philosophical points about the meaning of probability to gross logical errors such as claiming that large p-values are evidence that the null hypothesis is true (Greenland et al., 2016; Gigerenzer et al., 2004). Statisticians have produced a century's worth of clarifications, admonitions, and alternative approaches in hopes of correcting these mistakes (Neyman and Pearson, 1933; Hurlbert and Lombardi, 2009; Berger and Sellke, 1987; McShane et al., 2019; Amrhein et al., 2019; Harlow et al., 2013; Cohen, 1994a; Fidler et al., 2004; Schmidt and Hunter, 1997; Amrhein and Greenland, 2018; Bonovas and Piovani, 2023; Wasserstein and Lazar, 2016; Held and Ott, 2018). Many fields, including psychology (Henkel, 2017; Cohen, 1994a; Fidler et al., 2004; Harlow et al., 2013) and epidemiology (Goodman, 1993; Moher et al., 1999), have responded to these issues with decades of active debate and have reformed both how experiments are designed and how data is interpreted. In contrast, NHSTs have risen to near ubiquity in cell biology. For example, from 1980 to 2020, the number of *Journal of Neuroscience* papers using p-values increased from 35% to 98% (Calin-Jageman, 2022).

Among the various problems with how p-values are used, there is one convention that fundamentally undermines our ability to learn from data. Cell biologists define the test (null) hypothesis as the hypothesis that there was no effect (Gigerenzer, 2004). In the words of the statistician Jacob Cohen, the null hypothesis became the nil hypothesis (Cohen, 1994b). Most often, the no-effect hypothesis is the proposition that the control and experimental groups are random samples from the same population (control mean = experimental mean). There is no special reason, beyond convenience, to test the nil hypothesis as opposed to a hypothesis that there will be a specific difference between groups. However, this convenience has been a powerful force. Cell biologists have grown to believe that the purpose of data quantification is to answer the non-quantitative question of 'effect or no-effect'.

It may seem reasonable to ask whether an experimental treatment had an effect, but there is a major difference between the hallway definition of "no-effect" and the statistical definition of "no-effect". Imagine asking a colleague if the blood pressure drug they tested had an effect. Their experience will tell them that a reduction in blood pressure from 150 mmHg to 149.9 mmHg is not clinically relevant while a reduction to 125 mmHg could be life saving. If they excitedly tell you that the drug had an effect, you can assume the reduction wasn't by 0.1 mmHg. The same isn't true of a t-test. If you ask the t-test if the difference is zero, the p-value only tells you how often the observed difference in means -or more extreme differences- would occur if the true effect size was *exactly* zero.

**Rejecting the no-effect hypothesis tells us about the experiment, not the biology.**

In a highly interconnected network like a living organism, the proposition that one component is perfectly independent from another component is trivially false. By virtue of being part of the same organism, all molecular pathways and cells can be assumed to be either directly or indirectly connected. Detecting the connection might require extremely sensitive equipment and many samples, but the biological question is never "Is there a connection?" The meaningful question is always "How strong is the connection?".

The second problem with the no-effect hypothesis is that all experiments can be assumed to have some sampling bias (McShane et al., 2019). For instance, it is now recognized that circadian rhythms have detectable effects on most cellular processes. How much of the published biological literature has strong controls for time of day? Even if we could perfectly control the biology, there is no perfectly bias-free biology measuring device. Pointing out that no experiment is perfect is only necessary because no-effect tests are asking about an exact point value. Even if the true biological effect size were zero, we should not expect to record a non-zero experimental effect.

Finally, and most fundamentally, the statistical models underpinning NHSTs are mathematical simplifications of biological processes. The assumptions underlying these models are, at best, approximations of experimental conditions. Using these models to ask *how much bigger* the observed effect size is than the range of effects predicted for a given hypothesis is a great way to make sense of data (Resource box 1). However, there should be no expectation that any statistical model can perfectly predict potential experimental results (Amrhein et al., 2019).

The upshot of these three sources of guaranteed deviations from the nil hypothesis is that p-values for ALL no-effect tests will become infinitesimally small as sample sizes approach infinity. Imagine a dial that increases the sample size in all publications. As we turn the dial up, asterisks begin appearing over every plot and bar graph. We can keep turning the knob until each bar has a string of asterisks that runs off the page and the conclusion of every test is that the result was extremely statistically significant. The biology hasn't changed. The questions haven't changed. But increasing the sample size has guaranteed the same answer to every question.

**No-effect testing is imploding.**

In the past, it was common for cell biology experiments to consist of a small set of noisy measurements. When sample size is small and noise is high, a trivially small biological effect is unlikely to produce a small p-value. The difficulty of obtaining cell biological data also gave researchers a strong incentive to only perform experiments when they had a high expectation (high *a priori* probability) that an important biological effect would be revealed. Under these two conditions, low experimental sensitivity and high prior probability, most small p-values can be expected to correspond to non-trivial biological effects.

But what happens to no-effect testing when automation allows a massive amount of high-quality data to be acquired and analyzed with minimum human supervision?

1) Giant samples of trivial effects will produce arbitrarily small p-values.

2) We can automatically test thousands of variables that, individually, have low prior probabilities of being important. If follow-up experiments are evaluated with tests for no-effect, researchers will follow a drunkard's walk (Mlodinow, 2009) from one trivial true-positive to the next.

3) Automated analysis can generate many highly derived measures from the same data. Interpreting derived measures with tests for no-effect makes it possible to claim a result is significant without anyone understanding what is being measured.

**Effect size should be at the center of planning, analyzing, and discussing experiments.**

The reason testing for no-effect distorts scientific reasoning is that it allows us to analyze measurements without thinking about effect size. Effect size is usually measured as the difference in means between two groups, but effect size can refer to any quantifiable difference, relationship, or change. Effect size matters because: 1) Experimental measurements are about effect size. 2) Effect size determines the importance of interactions. 3) Models of cells run on effect sizes. 4) Effect sizes -unlike p-values- can be compared between experiments. 5) Consideration of effect size -unlike p-values- determines whether a field of inquiry is quantitative. Currently, it is difficult to argue that cell biology is operating as a quantitative science.

Prior to performing an experiment, we should be able to explain how a measure maps onto biological meaning. Ideally, we will have at least one real-world example of a small effect and one real-world example of an important effect. Imagine we have a mutant mouse model of a disease in which the number of mitochondria in each cell is reduced to half. We want to use this mouse model to test the efficacy of a drug treatment. The data we collect is the number of mitochondria, but we don't necessarily care whether a cell has 20 or 21 mitochondria. We care about recovery. We, therefore, define a scale with mutant mitochondria number at one end (0%) and healthy mitochondria number at the other end ($\%\,Recovery = \frac{Treatment - Mutant}{Healthy - Mutant} \times 100$, Figure 1A). When we talk about our results, we present both the raw mitochondria number and the percent recovery.

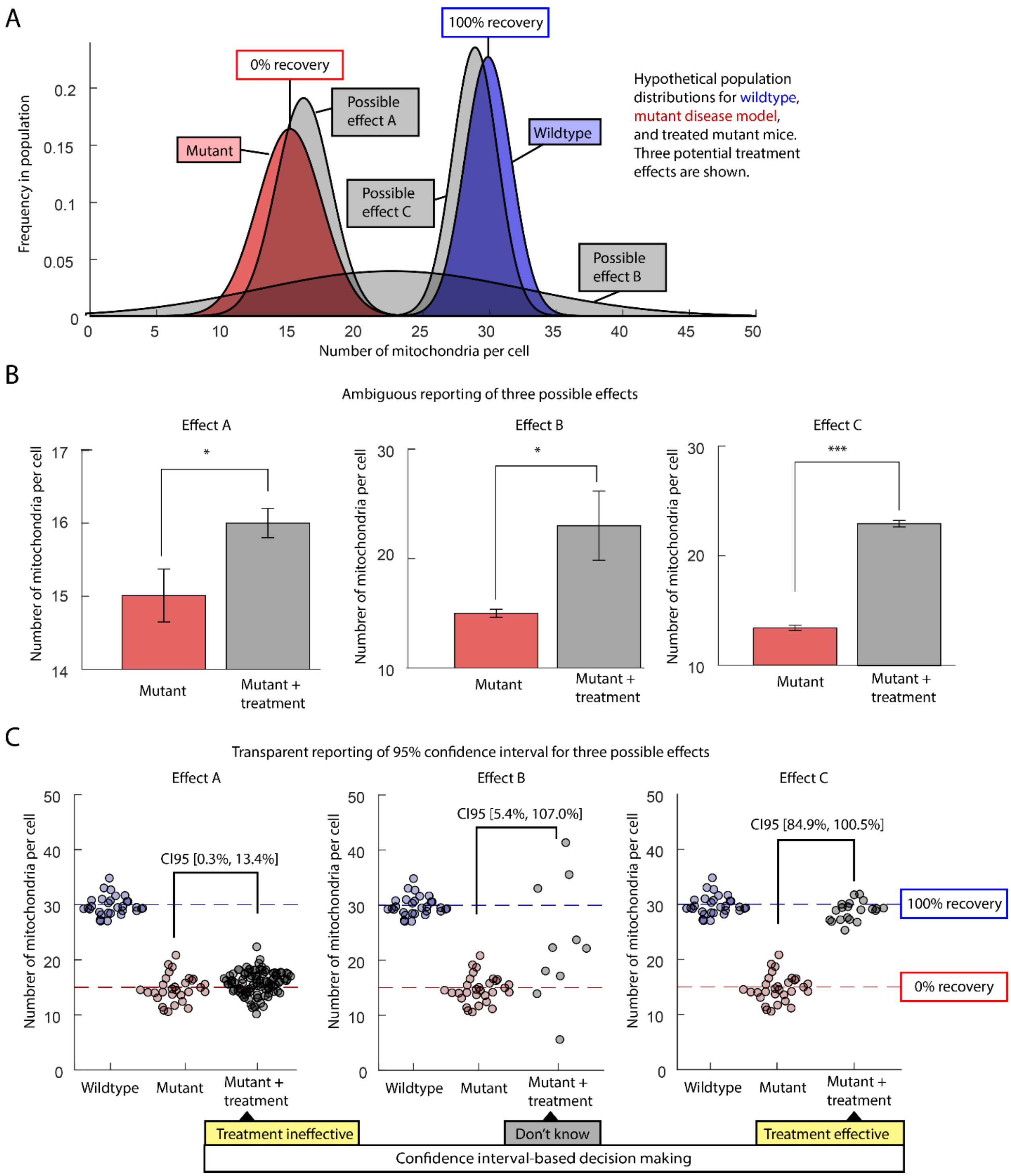

***Figure 1:*** *Comparison of hypothetical population distributions with different ways to represent results. Mitochondria number per cell is measured in healthy mice, a mutant mouse disease model, and in disease model mice that have undergone treatment. Three possible results are shown for the treatment. A) Hidden population distributions for mitochondria number in healthy (blue), disease model (red), and treatment (grey) mice. B) All treatment results are statistically significant when the no-effect hypothesis is tested. C) Scatterplots and confidence intervals are used to report the treatment effect (percent recovery). CI95 indicates the 95% confidence interval for the difference in*

*means between the mutant and mutant + treatment groups. The interval is converted to percent recovery by dividing the treatment effect by the difference between the mutant and wildtype (x100).*

How should we analyze the effects of a treatment tested on these mutant mice? Our worst option is to put off thinking about statistics until we have the data. We then perform the default t-test for no-effect. If the null hypothesis is rejected, we conclude that the treatment was effective. We report our results with a series of bar graphs and asterisks (Figure 1AB). By treating a weak version of a NHST as a rigorous test of a research hypothesis, we violate principles of science, statistics, and common sense. Critically, the p-value fails to distinguish between a trivial effect, an ambiguous result, and a complete cure (Figure 1AB). A reader who wants to determine if the treatment effect is biologically meaningful will have to decode the y axis of the plot, compare it to the standard error bars, then look up the sample size, and then re-read the text to try to determine what an important effect size might be.

NHSTs, weak or rigorous, are not a good way to understand biology. But, for the sake of argument, what would it take for a p-value to mean roughly what we would like it to mean? In this case, we need a quantitative definition of "ineffective treatment". Before the experiment, we might find that previous treatments achieved 5% - 10% recovery, a 30% recovery is required for a health improvement, and there is ~5% batch-to-batch variation in mitochondria counts. We could argue 20% recovery is, therefore, a reasonable cut off for identifying promising treatments. We would then need to define a p-value threshold that reflects the relative costs of following up a false positive result or neglecting to follow up on a false negative (Neyman and Pearson, 1933). We would analyze the variance of recovery measures to determine what kind of statistical model (parametric, non-parametric, non-linear) is appropriate. We should also perform a power analysis and preregister the experiment (Nosek et al., 2018). Finally, we would perform the experiment and perform a NHST for the hypothesis that the experimental effect was 20% or less. Given the totality of experimental design, data and analysis, a rejection of the null hypothesis would support the conclusion: "our best bet is to find out more about the treatment".

Is that really the best we can do? If we understand the biological meaning of potential experimental effects and we have designed a rigorous experiment for measuring those effects, why not just ask: "What was the effect?"

Reading data to understand effect size starts with staring at scatterplots (Figure 1C). Are the data points from each group clearly clustered around one point or are there multiple sub-clusters, extreme outliers, or a general lack of consensus? Are the data points spread across a biologically reasonable range of effects or are most of the data points bunched together near some detection threshold? If we plot rich experimental information, such which datapoints came from which experimental subject (see SuperPlots (Lord et al., 2020)), how much of our variance is coming from differences between measures, between individuals, or between batches? Finally, how big are the differences between groups compared to the differences within groups and the absolute value of our measures? There are worthwhile mathematical approaches to answering all these questions, but careful consideration of the scatterplot is a necessary sanity check and is often sufficient for making sense of a set of measures.

For thinking about the biology, reporting results, and making claims, we also want to calculate a conservative estimate of effect size. Effect size and precision are often summarized as a point estimate and a standard error (e.g., mean = 5.0 ± 0.2 SE). While these are useful statistics, they are not quite the quantification we need for evaluating the biological meaning of data. First, when we see a point value (mean, median, correlation coefficient) it is difficult for our brains not to register the point value as more 'biologically true' than a range of nearby values. Second, making good use of the standard error requires math, consideration of experimental design, and consideration of data characteristics.

When we report measurements, our goal should be to transparently report the range of effect sizes that make sense given the available data. This range is called an interval estimate. If we estimate that the recovery level of the treatment group was 84% to 100%, we can make a statistically and biologically meaningful claim that treated group looks more like the healthy mice than the disease model mice (Figure 1C). If our interval is limited to trivial effect sizes (0% to 13%), we can start looking for alternative treatments. If our effect size estimation includes a wide range of potential effects (5% to 107%), then we need to perform additional experiments or analyses to determine if the range of results is due to measurement error, biological variation, or treatment variation.

**Types of interval estimates**

Types of interval estimates include confidence intervals (frequentist statistic)(Neyman, 1937), credible intervals (Bayesian statistic)(Morey et al., 2016), prediction intervals (forecasting future results)(Billheimer, 2019), tolerance intervals (defined by proportion of population)(Owen, 1964), and likelihood intervals (likelihood function)(Hudson, 1971). It is also possible to report a two-sided (50% to 103%) or one-sided interval (greater than 60%). Choosing the best method estimating an effect size interval requires careful consideration of how much is already known about the system and the goals of the research program (Hahn and Meeker, 1991).

In the succeeding discussion, I will focus almost exclusively on two-sided confidence intervals and I will recommend that confidence intervals be adopted as the next default statistic in cell biology. This recommendation is not because confidence intervals are ideal for every cell biology experiment. The recommendation reflects the current culture of cell biology and how we do experiments. Key to this recommendation is the fact that our field can make the profound switch to effect-size-based science simply by reporting the confidence intervals generated by the same statistical models we previously used to perform NHSTs.

**CI95% should be cell biology's next default statistic**

NHST for no-effect became ubiquitous because there is a genuine niche for a simple and universally understood quantification of what we see when we look at scatterplots. Reporting p-values was a reasonable idea, but overuse, misuse and misinterpretations proved disastrous. Compared to p-values, confidence intervals are an efficient and transparent approach to reporting both effect size and precision (Resource text 2).

Confidence intervals also respond in an intuitive way to how much information is available. More data means, on average, greater precision of effect size estimation. As a default statistic, confidence intervals have a huge advantage in that they provide useful information across wide ranges of experimental designs including preliminary experiments, rigorous descriptive studies, qualitative hypothesis testing, and quantitative hypothesis testing. Given these benefits, confidence intervals are the most commonly proposed alternative to p-values (Cohen, 1994b; Thompson, 2014; Neyman, 1937; Nakagawa and Cuthill, 2007).

Treating confidence intervals as cell biology's default statistic (technically our default inferential construct) does not mean that all data must be analyzed with a confidence interval. Making it our default statistic means that the confidence interval is what we would use when we collect measurements and don't have a better idea of how to summarize our data. We would make sure that every student is comfortable calculating and interpreting a confidence interval. When readers, reviewers, and editors evaluate a publication, the confidence interval would define the minimum level of statistical sophistication and meaning that they should expect from the analysis of sample measurements.

**What does the 95% confidence interval mean?**

There are a variety of parametric and non-parametric methods for calculating a 95% confidence interval (Resource Box 3), but the goal is the same. The *nominal* performance of a 95% confidence interval means that, in an infinite number of ideal experiments, the real test parameter would fall within the bounds of the interval in 95% of trials (Figure 2AB). How close the *actual* interval performance gets to the nominal performance depends on how well the assumptions of the model match the conditions of the experiment (Miao and Chiou, 2008; Puth et al., 2015). Critically, the '95%' value is NOT the probability that an individual interval includes the true population value. A single interval is either right or wrong.

What, then, does a CI95% mean in practice? If the statistical model is appropriate to the experiment and the experiment is well constructed, it is reasonable for us to think of the biological effect as existing within the bounds of the 95% confidence interval. We also need to understand that we will be wrong more than 5% of the time. Accepting that at least 5% of cell biology experiments will produce misleading results should not rock our view of the field (Border et al., 2019; Errington et al., 2021). Our actual confidence in our understanding of a system should reflect the accumulation of consistent evidence from independent experiments that, ideally, use independent experimental approaches. In some cases, this accumulation of evidence can be quantified (Santos et al., 2024). Most of the time, we will have to rely on our own critical evaluation of the literature (Nosek and Errington, 2020).

One advantage to using confidence intervals is that they make sense even when experiments are preliminary or exploratory. Calculating a confidence interval does not require a hypothesis or prior information about effect size or variance size. Calculating a confidence interval does require a model of variance and a minimum number of samples. However, using a t-distribution (ttest) to calculate a confidence interval from ten data points that look Gaussian will usually get us in the right ball park (Figure 2A-D). Note that confidence intervals merge estimates of effect size with estimates of variance. At small sample sizes (<8), the estimates of variance degrade and the interval widths vary wildly (dashed curves, Figure 2C). At these sample sizes, reporting a mean and an a priori precision (Trafimow, 2017) (Figure 2C, red line, see below) might be a more useful way to discuss results.

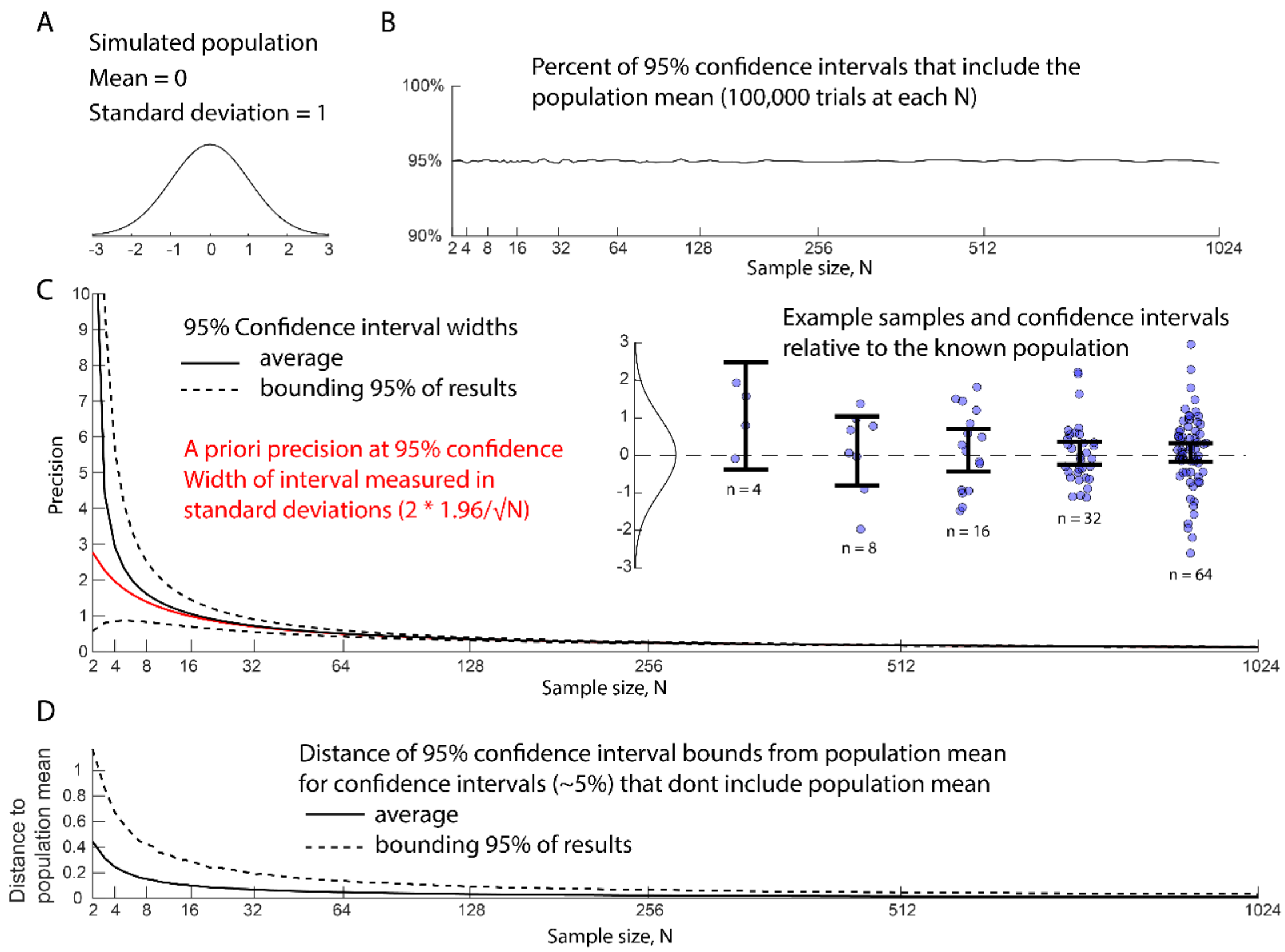

*__Figure 2.__ Performance of a Welch-Satterthwaite 95% confidence interval when estimating the mean of a simulated normal distribution. Each sample size was tested with 100,000 samplings from the simulated population. A) Simulated population parameters. B) Percent of 95% confidence intervals that include the known population mean. The interval performs close to the nominal value. C) Width of confidence interval (precision) relative to sample size. Black line indicates the average width of the confidence interval. Dotted lines bound confidence interval width of 95% of trials. Red line shows the 95% a priori precision displayed as fractions of standard deviations. Inset shows example trials and confidence intervals relative to the sampled population. D) The confidence intervals that don't include the population mean (~5%) still approach the population mean as sample size increases. Traces are for trials where the confidence interval did not include the simulated population mean. The black line is the average distance of the population mean to the nearest boundary of the confidence interval. The dotted line is the distance between the population mean and confidence interval that includes 95% of trials.*

The most important function of confidence intervals is that they can be used to rigorously estimate effect sizes. If we want to know how a mutation in protein A relates to the rate of cell division, we can measure the rate of cell division in animals with and without the mutation. We can choose a sample size that reflects the effect sizes we are interested in and the known variance of the rates. We can then calculate a confidence interval for how much the mutation changed the rate of cell division (CI95[55%,59%]). That interval can help us understand clinical impacts of the mutation, build models of the regulation of cell division, or compare the effect of protein A to the effects of other possible regulators of cell division.

It is also possible to test a hypothesis with a confidence interval. What if our working model of protein A predicts the mutation would increase the rate of cell division by 90%? A convenient symmetry to confidence intervals is that values outside of their bounds can be treated as hypotheses that are not compatible with the data (Amrhein and Greenland, 2022). In the case of the 90% prediction and the CI95[55%,59%] result, we would reject the test hypothesis. More importantly, the confidence interval shows how far the null hypothesis is from the estimated range of effect sizes. A confidence interval whose closest bound is far from the null hypothesis argues that there is something important going on that is not adequately described by the null hypothesis. A narrow confidence interval surrounding the test hypothesis would be a failure to reject the working model. A wide confidence interval that includes the test hypothesis could be interpreted as too weak an experiment to affect our view of the model. Note that the distinction between "far from the null hypothesis" and "close to the null hypothesis" depends on the biologist's understanding of the measurement.

**Do we need to falsify a hypothesis?**

For many years, cell biologists were told that good science is hypothesis-driven science and that hypothesis-driven science means performing manipulation experiments that either reject or fail to reject a mechanistic hypothesis. Studies that did not fit this formula were often dismissed as "merely descriptive". Analyzing every experiment with a NHST to reject a no-effect strawman was treated as consistent with the Popperian program of hypothesis falsification.

It is difficult to summarize all the problems with the above view of science. Rejecting an impossible hypothesis isn't falsification science. But falsification science is also not how cell biology works. To the philosopher Karl Popper, the question of where theories came from was "irrelevant to the logical analysis of scientific knowledge" (Popper, 1959). To cell biologists, theories come from centuries of hard work building instruments that can observe things that couldn't be observed before and then systematically doing the observing. Rigorous non-hypothesis-testing studies are the bulk of the research that needs to be funded and published if we want our field to generate theories that are worth testing.

Cell biology has produced an incredibly successful, unified and coherent model of how cells evolve, develop, behave, interact, build tissues, and fall apart. When we identify important gaps in this model, we map and manipulate the phenomena of interest until the gap is filled. If the next big gap in knowledge is the sequence of the human genome, and we can map the human genome, then no hypothesis middleman needs to get involved. This logic includes experimental manipulations. If there is a good reason to care about how much a mutation increases cell division, and we can measure that increase, framing our analysis as a hypothesis test is misleading and unnecessary.

**Some limits of confidence intervals**

The difference between a p-value and a confidence interval is in which part of a statistical prediction is binarized. You can report a p-value, the precise fraction of a probability distribution defined by two magnitudes (null hypothesis and observed result). Alternatively, you can report a confidence interval, two magnitudes corresponding to a defined fraction of a probability distribution. In favor of reporting confidence intervals is the fact that there is no clear way to think about what a precise fraction of a hypothetical probability distribution means for our understanding of the world. The widespread confusion about p-values and probability reflects not just a lack of statistical sophistication, but also a genuine epistemological conundrum. In contrast, the values reported in a confidence interval are numbers with clear biological interpretations.

It is possible to visualize the full range of a model's predictions without imposing the threshold required for a p-value or confidence interval. A compatibility curve is a two-dimensional plot of the p-values (y-axis) that would be obtained for a range of plausible hypotheses (x-axis) given the observed data and a statistical model (Amrhein and Greenland, 2022; Berner and Amrhein, 2022; Poole, 1987) (Figure 3). The example shows the compatibility curves generated by two simulated experiments (n=20) that estimate a population mean using t-distributions (a bunch of t-tests). The two points where the compatibility map intersects p = 0.05 are equivalent to the bounds of the 95% confidence interval. Note, that the peak of each compatibility curve (p = 1) is simply the mean of the experimental sample. There is no statistical magic that suggests that these peaks are “probably” the true mean. Reporting the results of a statistical model with a compatibility curve is especially useful for non-normal distributions that are not readily summarized by confidence intervals.

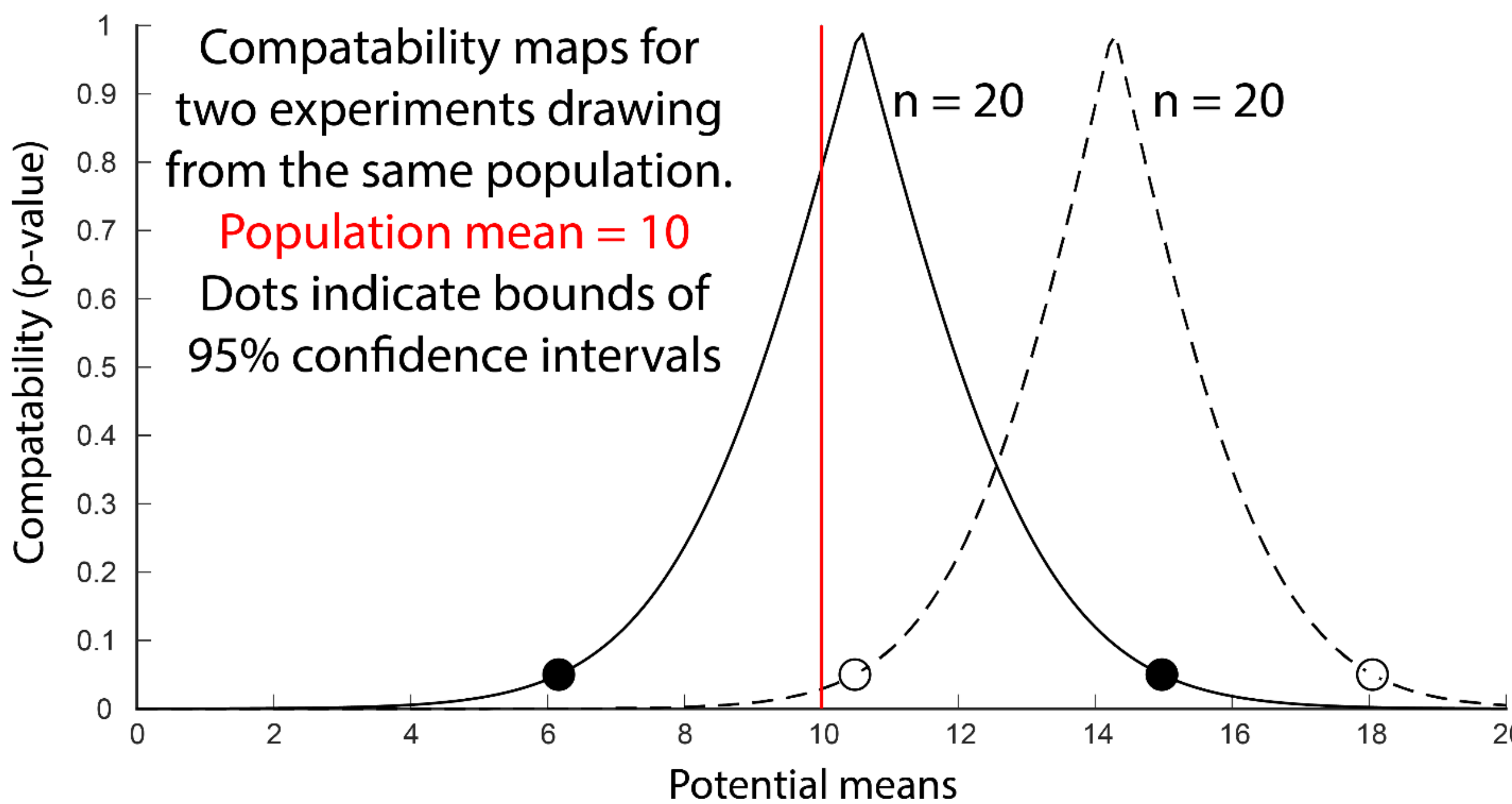

***Figure 3.*** *Compatibility maps produced by two simulated experiments (solid line and dotted line) that sampled from the same population. Each trace shows the p-value (y-axis) that would be produced by testing the experimental results against a range hypotheses (x-axis) about the population mean. The simulated population had a mean of 10 (red line) and a standard deviation of 10. The 95% confidence intervals are indicated by the dots.*

The close relationship between confidence intervals and p-values means that confidence intervals are subject to many of the same limits and misinterpretations as p-values (Trafimow et al., 2024; Morey et al., 2016). If confidence intervals become the default summary statistic in cell biology, we can expect to see similar overinterpretations of what confidence intervals can tell us about our data (Greenland et al., 2016). We can also expect effect sizes to be systematically inflated by measurement bias (Vul et al., 2009; Fiedler, 2011), publication bias (Yang et al., 2023), and the winner’s curse (Button et al., 2013; Ioannidis, 2008).

One form of measurement bias warrants special attention because it often results from an attempt to objectively analyze large and complicated data sets (images and ‘omics). A researcher will perform an experiment with thousands of potentially interesting variables (voxels, RNAs, genes, or cells). They use a criterion such as a p-value for identifying which variables were affected by some manipulation. They then calculate an effect size for the identified variables. The problem is that the variance that influence selection is not independent from the variance that influences effect size estimation. Ironically, the more conservative the selection criterion ($p < 0.00001$), the more the average reported effect size will be inflated (Ioannidis, 2008). The solutions include defining an independent criterion for selection (Fiedler, 2011) or performing an adjustment to the effect size estimate that takes the selection criterion into account (Holland et al., 2016).

While confidence intervals are a good starting point for cell biology, they do not make full use of the available data. As a measure of precision, confidences intervals lump together different classes of error (measurement precision, variation within the population, and sampling error) that could be calculated and reported separately (Trafimow, 2018). If the data exists for positing prior probabilities, then a Bayesian credibility interval will leverage more of the available information than a confidence interval (Morey et al., 2016). More generally, statisticians point out that statistical techniques are constantly being improved and that it is wrong headed to expect one type of analysis to be optimal, or even appropriate, for most experiments (Wasserstein and Lazar, 2016; Amrhein and Greenland, 2018; Cumming and Calin-Jageman, 2016).

Given the limitations of confidence intervals, why is it worth pushing tens of thousands of cell biologists to replace ‘p-value ’ with ‘CI95’ ? Currently, cell biologists can publish sample measurements without discussing effect size. Asking cell biologists to stop what they are doing and read an effect-size-centric statistics text book (Cumming and Calin-Jageman, 2016) before designing their next experiment is not realistic or polite. Switching to confidence intervals, in contrast, provides useful information about effect size, requires no change in experimental design, requires no change in statistical model, and makes p-values superfluous. Confidence intervals are a remarkably low-cost first-step towards overturning a dysfunctional status quo.

**Alternatives to the philosophical confusion of confidence intervals**

As humans and scientists, we theorize there is a universe that we can collect information about through our experiences. From these experiences, we build mental models of bits of the universe (inference) that make predictions about future experiences. If the universe works like X, we can expect Y to happen. A statistical model is an abstracted version of a cognitive model whose behavior can be described mathematically. The brilliance of these statistical models is that they can be applied by many different observers, to many different processes, and that their predictions are repeatable and quantifiable. However, the extent to which the model tells you something useful about the universe depends on how well each assumption of the model corresponds the behavior of the patch of the universe you are investigating.

Our statistical claims are philosophically valid when we treat statistical inference as thought experiments instead of facts about the world (Amrhein et al., 2019). However, a useful cognitive model requires that we eventually summarize evidence as something like: "I'm pretty sure most cell bodies are about 10 μm in diameter". The numbers from statistical analysis can help us refine 'pretty sure', 'most, and 'about', but the relationship isn't one-to-one. That is, the probability distributions predicted by a statistical model exist within the bounds of our uncertainties about both the appropriateness of the model and reliability of the experimental controls.

One path out of these philosophical ambiguities is to perform our analysis of precision before conducting the experiment. Trafimow and colleagues argue for combining an a priori procedure (APP) for calculating precision with a gain-of-probability (G-P) diagram to analyze results (Trafimow et al., 2024; Trafimow, 2017). The simplest version of APP is to determine the sample size needed to estimate a mean. We first define our target precision as some number of standard deviations between a true mean and a measured mean (±0.5 SD). We then define our target probability that a measured mean would fall within this precision range (for 95%, z-score = 1.96). Finally, we calculate the minimum sample size that would satisfy our two goals ($n = \left(\frac{1.96}{.5}\right)^2 = 16$).

A second path out of the philosophical confusion is for more cell biologists to perform statistical analysis by programming Monte Carlo simulations (Buckland, 1984)(Resource Box 1). Programming these simulations allows a researcher who is unfamiliar with the fine points of statistical theory to ask their question in the most straightforward way possible: "If the system worked like this and I ran my experiment 100,000 times, what would I expect to see?" The act of writing the simulation forces the researcher to think of the statistical model as a system of "if, then" statements that may or may not match reality. It can also push the researcher to question whether the hypothesis they are programing is really a plausible way for the system to work.

**What happens to cell biology if we stop reporting tests for "no-effect"?**

The criteria of "biologically meaningful effect size" is a much higher bar than "statistically significant". Under this criterion, many fewer studies can be expected to claim important experimental effects. This aspect of increased statistical rigor does not necessarily mean fewer papers will be published. An unbiased literature requires publishing both "positive" and “negative” results. Focusing our analysis on effect size does mean that we will have to accept the reality that most experimental results are ambiguous and incremental. The implicit and, sometimes, explicit interpretation of p-values is that the result of every experiment is either “effect” or “no-effect”. When we take interval estimates seriously, we will often have to conclude that we don’t have enough data to understand what is going on.

Discussing confidence intervals will also require more work than discussing tests for no-effect. It is convenient for everyone to agree that there is a statistic (p-value) that tells you that something important happened even when it is not clear what is being measured. In the absence of this fiction, our claims will require more explanation about how our measures relate to the biology and how the observed effect sizes relate to our question.

Arguing that we should produce a more detail-oriented literature filled with ambiguous and negligible effects is not the most appealing pitch for changing the way cell biology quantifies data. What we could gain is a literature in which biological claims reflect meaningful quantification, a literature from which we can judge the replicability of experiments, and a literature from which we can build better models of how cells work.

Imagine again the dial that increases sample size in all publications. This time, the publications base their claims on confidence intervals. As sample sizes increase towards infinity, the intervals shrink to point values that are the actual value of the population. When effect size estimation is the foundation of our analysis, more data translates into more understanding.

## TEXT BOXES

### Comparing an experimental effect to a Monte Carlo distribution

Imagine we want to know if synaptic inputs to a neuron are concentrated on the apical dendrite of some type of neuron. We perform an experiment where we find that the densities of synapses on the apical dendrites are about twice as high as the densities on the rest of the cell. What we want to know is if this discrepancy can be explained by how often the apical dendrite encounters potential synaptic partners. We can build a Monte Carlo simulation of the tissue that includes anatomical constraints such as the geometry of the cells and the distribution of cells in space. Within the model, a fixed number of synapses are distributed by identifying places where axons and dendrites are close to one another. We run the model 100,000 times, each time varying the positions of cells and recording the resulting density of synapses on apical dendrites. How do we compare our observed results to the Monte Carlo model?

We could generate a p-value by counting how many of the results have an equal or greater density of synapses on the apical dendrites. A very low p-value might support the argument that the observed bias would be unlikely to be produced by the anatomy as simulated. But what if we forgot to factor in subtle changes in the diameters of dendrites? What if the density of blood vessels is different for different regions of the tissue and they obstruct some opportunities for making synapses? A small p-value could also result from any number of trivial anatomical biases that were not controlled for in the model.

The question we should care about is how much of the bias in synapse density can be explained by a reasonable set of anatomical constraints. To make this comparison, we calculate the range of values that encompass 95% of the Monte Carlo results. We can then compare these bounds to our observed bias. One possible result is that the density of synapses we observe on the apical dendrite (400/mm) is far outside of the predicted range (CI95[150/mm, 250/mm], Resource Figure). In this case, we can now use the upper bound of the Monte Carlo range to report a conservative estimate of the bias towards forming synapses on the apical dendrite: "The observed apical dendrite density was at 60% higher than the upper bound predicted by our model". We might also find that our observed density is within the range predicted by our model. In that case, it is useful to report if inclusion of the observed value is because the model precisely predicts the density (CI95[390/mm, 410/mm]) or because the model predictions include a wide range of outcomes (CI95[50/mm, 800/mm]). Regardless of the result, a histogram of the experimental results plotted against the Monte Carlo results provides the most transparent description of the analysis (Resource Figure).

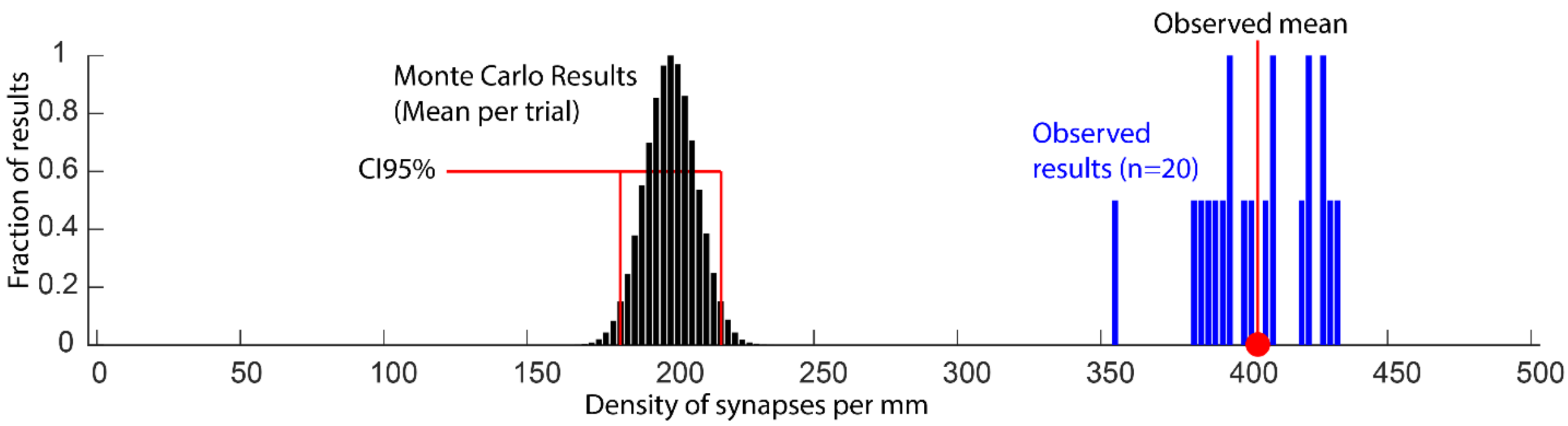

***Resource Figure.*** *Comparison of observed results to the results of a Monte Carlo simulation. The mean synaptic density of apical dendrites from each of 100,000 Monte Carlo simulations are shown in black. The bounds of the 95% confidence interval encompassing these results are bracketed in red. The observed sample is shown in blue with the mean highlighted with the red circle. The conservative estimate of effect size is the distance between the red circle and the nearest CI95 bracket.*

**Calculating confidence intervals.**

There are many ways to calculate confidence intervals. When the t-distribution of a ttest is appropriate for calculating a p-value, the t-distribution is also appropriate for calculating a confidence interval. That is why the easiest path to effect-size-centric cell biology is to run the same ttests we normally would perform but report the CI95 instead of $p < 0.05$. When a non-parametric bootstrap test would be the best way to calculate a p-value, then a bootstrap confidence interval will probably be the best approach (Efron, 1979; Fieberg et al., 2020; Jung et al., 2019; Shi, 1992; Letson and McCullough, 1998; Puth et al., 2015; DiCiccio et al., 1992).

To pick the best CI95 for an experiment, we need to consider both the assumptions of the statistical model (sample size, normality, etc.) and the performance of the model. A confidence interval that performs well for our data type will produce robust, stable, and narrow intervals whose actual coverage is close to the nominal (95%) coverage. 'Robust' means that it can tolerate deviations from the ideal experimental conditions (normal data vs. normal-ish data). 'Stable' means that the width of the interval is consistent between samples from the same population. The performance of common methods of calculating confidence intervals have been studied systematically (Puth et al., 2015; Miao and Chiou, 2008). However, generating simulated data yourself and experimenting with different methods of calculating intervals is worthwhile both as a learning experience and as a path to finding the best way to analyze unconventional data (https://github.com/MorganLabShare/betterThanChance).

If we decide that a t-distribution is a good model for our data, we still have to decide how to perform the calculation. What if we want to know if a published claim about the difference between two groups would still make sense if a confidence interval had been calculated? If we can pull a difference in means and standard error from the text or a bar graph, we can calculate a simple back-of-the-envelope confidence interval. We can combine the standard errors of each group to calculate a pooled standard error (pooled SE = sqrt(SE1^2 + SE2^2). We can calculate the margin of error by multiplying the pooled standard error with the t-score (1.96 for 95% interval). Adding and subtracting the margin of error to the published difference in means gives us a CI95. While this calculation is convenient when no computer is at hand, it performs poorly when the sample size is small or when sample size differs between groups.

The Welch-Scatterhwaite (WS) confidence interval is much more robust because of how it incorporates group differences into its calculation of pooled error and degrees of freedom (Welch, 1938). Miao and Chiou (Miao and Chiou, 2008) show that it performs well when sample sizes are the same or different, when variances are the same or different, and even when distributions are not normal. They do find that its performance decreases when distributions are asymmetric. They recommend first testing the data for symmetry and then performing a log transformation when the data is asymmetric. With the caveat of detecting and correcting for asymmetric distributions, the WS confidence interval is a single calculation that could reliably replace most of the p-values used in cell biology.

WS confidence intervals can be generated using standard in statistics packages and it is the default calculation used for t-tests in R (R Foundation) and Matlab (Mathworks). In R, the WS CI95 is included in the result of: t.test(group1, group2). In Matlab 2023a, the WS CI95 can be obtained by the expression: [noThanks1 noThanks2 wsCI95] = ttest2(group1, group2, "Alpha", 0.05). To calculate the interval by hand:

Welsh-Scatterhwaite confidence interval for a difference in means

1. Calculate the combined standard error (*sp*) using the standard deviation (*std1, std2*) and sample size (*n1, n2*) for two groups: $w1 = \frac{std1^2}{n1}, w2 = \frac{std2^2}{n2}, sp = \sqrt{(w1} + w2)$
2. Calculate the degrees of freedom (*df*): $df = (w1 + w2)^2/(\frac{w1^2}{n1-1} + \frac{w2^2}{n2-1})$

3. Look up the t-score (*t*) for the desired alpha using the t-distribution with *df* degrees of freedom. For a 95% confidence interval, alpha = 0.05: $t = tDistribution([1 - \frac{alpha}{2}], df)$
4. Calculate the margin of error (*ME*): $ME = df * t$
5. Add and subtract the margin of error to the difference in means: $CI95 = (\bar{x}1 - \bar{x}2) \pm ME$

## CODE AVAILABILITY

The Matlab (Mathworks) code used to generate figures and that can be used to experiment with different types of distributions is available at https://github.com/MorganLabShare/betterThanChance.

## ACKNOWLEGEMENTS

Thanks to Hrvoje Šikić, Tim Holy, Phil Williams, Daniel Kerschensteiner, and Julie Hodges for reading the manuscript and providing feedback. This work was supported by an unrestricted grant to the Department of Ophthalmology and Visual Sciences from Research to Prevent Blindness, by a Research to Prevent Blindness Career Development Award, and by the NIH (EY029313).

## REFERENCES

Amrhein, V., and S. Greenland. 2018. Remove, rather than redefine, statistical significance. *Nat Hum Behav*. 2:4.

Amrhein, V., and S. Greenland. 2022. Discuss practical importance of results based on interval estimates and *p*-value functions, not only on point estimates and null *p*-values. *J. Inf. Technol.* 37:316–320.

Amrhein, V., D. Trafimow, and S. Greenland. 2019. Inferential statistics as descriptive statistics: There is no replication crisis if we don't expect replication. *Am. Stat.* 73:262–270.

Bakan, D. 1966. The test of significance in psychological research. *Psychol. Bull.* 66:423–437.

Berger, J.O., and T. Sellke. 1987. Testing a Point Null Hypothesis: The Irreconcilability of P Values and Evidence. *J. Am. Stat. Assoc.* 82:112–122.

Berkson, J. 1938. Some Difficulties of Interpretation Encountered in the Application of the Chi-Square Test. *J. Am. Stat. Assoc.* 33:526–536.

Berner, D., and V. Amrhein. 2022. Why and how we should join the shift from significance testing to estimation. *J. Evol. Biol.* 35:777–787.

Billheimer, D. 2019. Predictive inference and scientific reproducibility. *Am. Stat.* 73:291–295.

Bonovas, S., and D. Piovani. 2023. On p-Values and Statistical Significance. *J. Clin. Med. Res.* 12. doi:10.3390/jcm12030900.

Border, R., E.C. Johnson, L.M. Evans, A. Smolen, N. Berley, P.F. Sullivan, and M.C. Keller. 2019. No support for historical candidate gene or candidate gene-by-interaction hypotheses for major depression across multiple large samples. *Am. J. Psychiatry*. 176:376–387.

Buckland, S.T. 1984. Monte Carlo Confidence Intervals. *Biometrics*. 40:811.

Button, K.S., J.P.A. Ioannidis, C. Mokrysz, B.A. Nosek, J. Flint, E.S.J. Robinson, and M.R. Munafò. 2013. Power failure: why small sample size undermines the reliability of neuroscience. *Nat. Rev. Neurosci.* 14:365–376.

Calin-Jageman, R.J. 2022. Better Inference in Neuroscience: Test Less, Estimate More. *J. Neurosci.* 42:8427–8431.

Cohen, J. 1994a. The earth is round (p < .05). *Am. Psychol.* 49:997–1003.

Cohen, J. 1994b. The earth is round (p < 05). *Am. Psychol.* 49:997–1003.

Cumming, G., and R. Calin-Jageman. 2016. Introduction to the new statistics: Estimation, open science, and beyond. Routledge, London, England. 564 pp.

DiCiccio, T.J., M.A. Martin, and G.A. Young. 1992. Fast and accurate approximate double bootstrap confidence intervals. *Biometrika*. 79:285–295.

Efron, B. 1979. Bootstrap Methods: Another Look at the Jackknife. *aos*. 7:1–26.

Errington, T.M., M. Mathur, C.K. Soderberg, A. Denis, N. Perfito, E. Iorns, and B.A. Nosek. 2021. Investigating the replicability of preclinical cancer biology. *Elife*. 10. doi:10.7554/eLife.71601.

Fidler, F., N. Thomason, G. Cumming, S. Finch, and J. Leeman. 2004. Editors can lead researchers to confidence intervals, but can't make them think: statistical reform lessons from medicine. *Psychol. Sci.* 15:119–126.

Fieberg, J.R., K. Vitense, and D.H. Johnson. 2020. Resampling-based methods for biologists. *PeerJ*. 8:e9089.

Fiedler, K. 2011. Voodoo correlations are everywhere-not only in neuroscience. *Perspect. Psychol. Sci.* 6:163–171.

Gigerenzer, G. 2004. Mindless statistics. *J. Socio Econ.* 33:587–606.

Gigerenzer, G., S. Krauss, and O. Vitouch. 2004. Th e null ritual what you always wanted to know about signifi cance testing but were afraid to ask.

Goodman, S.N. 1993. p values, hypothesis tests, and likelihood: implications for epidemiology of a neglected historical debate. *Am. J. Epidemiol.* 137:485–96; discussion 497-501.

Greenland, S., S.J. Senn, K.J. Rothman, J.B. Carlin, C. Poole, S.N. Goodman, and D.G. Altman. 2016. Statistical tests, P values, confidence intervals, and power: a guide to misinterpretations. *Eur. J. Epidemiol.* 31:337–350.

Hahn, G.J., and W.Q. Meeker. 1991. Statistical intervals: A guide for practitioners. John Wiley & Sons, Nashville, TN. 416 pp.

Harlow, L.L., S.A. Mulaik, and J.H. Steiger. 2013. What if there were no significance tests?

Held, L., and M. Ott. 2018. On P-Values and Bayes Factors. doi:10.1146/annurev-statistics-031017-100307.

Henkel, R.E. 2017. The Significance Test Controversy: A Reader. Routledge. 347 pp.

Hogben. 1957. Statistical Theory: The Relationship of Probability, Credibility and Error : an Examination of the Contemporary Crisis in Statistical Theory from a Behaviourist Viewpoint. Norton. 510 pp.

Holland, D., Y. Wang, W.K. Thompson, A. Schork, C.-H. Chen, M.-T. Lo, A. Witoelar, Schizophrenia Working Group of the Psychiatric Genomics Consortium, Enhancing Neuro Imaging Genetics through Meta Analysis Consortium, T. Werge, M. O'Donovan, O.A. Andreassen, and A.M. Dale. 2016. Estimating effect sizes and expected replication probabilities from GWAS summary statistics. *Front. Genet.* 7:15.

Hudson, D. 1971. Interval estimation from the likelihood function. *Journal of the royal statistical society series b-methodological*. 33:256–262.

Hurlbert, S.H., and C.M. Lombardi. 2009. Final Collapse of the Neyman-Pearson Decision Theoretic Framework and Rise of the neoFisherian. *anzf*. 46:311–349.

Ioannidis, J.P.A. 2008. Why most discovered true associations are inflated. *Epidemiology*. 19:640–648.

Jung, K., J. Lee, V. Gupta, and G. Cho. 2019. Comparison of Bootstrap Confidence Interval Methods for GSCA Using a Monte Carlo Simulation. *Front. Psychol.* 10:2215.

Letson, D., and B.D. McCullough. 1998. Better Confidence Intervals: The Double Bootstrap with No Pivot. *Am. J. Agric. Econ.* 80:552–559.

Lord, S.J., K.B. Velle, R.D. Mullins, and L.K. Fritz-Laylin. 2020. SuperPlots: Communicating reproducibility and variability in cell biology. *J. Cell Biol.* 219. doi:10.1083/jcb.202001064.

McShane, B.B., D. Gal, A. Gelman, C. Robert, and J.L. Tackett. 2019. Abandon Statistical Significance. *Am. Stat.* 73:235–245.

Meehl, P.E. 1967. Theory-testing in psychology and physics: A methodological paradox. *Philos. Sci.* 34:103–115.

Miao, W., and P. Chiou. 2008. Confidence intervals for the difference between two means. *Comput. Stat. Data Anal.* 52:2238–2248.

Mlodinow, L. 2009. The drunkard's walk: How randomness rules our lives. Random House, New York, NY. 252 pp.

Moher, D., D.J. Cook, S. Eastwood, I. Olkin, D. Rennie, and D.F. Stroup. 1999. Improving the quality of reports of meta-analyses of randomised controlled trials: the QUOROM statement. *Lancet*. 354:1896–1900.

Morey, R.D., R. Hoekstra, J.N. Rouder, M.D. Lee, and E.-J. Wagenmakers. 2016. The fallacy

of placing confidence in confidence intervals. *Psychon. Bull. Rev.* 23:103–123.

Morrison, D.E.H.R.E. ed. . 1970. The significance test controversy. Aldine, Chicago.

Nakagawa, S., and I.C. Cuthill. 2007. Effect size, confidence interval and statistical significance: a practical guide for biologists. *Biol. Rev. Camb. Philos. Soc.* 82:591–605.

Neyman, J. 1937. Outline of a theory of statistical estimation based on the classical theory of probability. *Philos. Trans. R. Soc. Lond.* 236:333–380.

Neyman, J., and E.S. Pearson. 1933. On the problems of the most efficient tests of statistical hypotheses. *Philosophical Transactions of the Royal Society of London*. 231A:289–338.

Nosek, B.A., C.R. Ebersole, A.C. DeHaven, and D.T. Mellor. 2018. The preregistration revolution. *Proc. Natl. Acad. Sci. U. S. A.* 115:2600–2606.

Nosek, B.A., and T.M. Errington. 2020. What is replication? *PLoS Biol.* 18:e3000691.

Owen, D.B. 1964. Control of percentages in both tails of the normal distribution. *Technometrics*. 6:377–387.

Poole, C. 1987. Beyond the confidence interval. *Am. J. Public Health*. 77:195–199.

Popper, K.R. 1959. The Logic of Scientific Discovery. *Hutchinson, London*.

Puth, M.-T., M. Neuhäuser, and G.D. Ruxton. 2015. On the variety of methods for calculating confidence intervals by bootstrapping. *J. Anim. Ecol.* 84:892–897.

Rozeboom, W.W. 1960. The fallacy of the null-hypothesis significance test. *Psychol. Bull.* 57:416–428.

Santos, J.A.R., R. Grant, and G.L. Di Tanna. 2024. Bayesian meta-analysis of health state utility values: A tutorial with a practical application in heart failure. *Pharmacoeconomics*. 42:721–735.

Schmidt, F.L., and J.E. Hunter. 1997. What if there were no significance tests? 37–64.

Shi, S.G. 1992. Accurate and efficient double-bootstrap confidence limit method. *Comput. Stat. Data Anal.* 13:21–32.

Sterling, T.D. 1959. Publication Decisions and their Possible Effects on Inferences Drawn from Tests of Significance—or Vice Versa. *J. Am. Stat. Assoc.* 54:30–34.

Szucs, D., and J.P.A. Ioannidis. 2017. When Null Hypothesis Significance Testing Is Unsuitable for Research: A Reassessment. *Front. Hum. Neurosci.* 11:390.

Thompson, B. 2014. The Use of Statistical Significance Tests in Research. *J. Exp. Educ.* 61:361–377.

Trafimow, D. 2017. Using the coefficient of confidence to make the philosophical switch from A posteriori to A Priori inferential statistics. *Educ. Psychol. Meas.* 77:831–854.

Trafimow, D. 2018. Confidence intervals, precision and confounding. *New Ideas Psychol.* 50:48–53.

Trafimow, D., T. Tong, T. Wang, S.T.B. Choy, L. Hu, X. Chen, C. Wang, and Z. Wang. 2024. Improving inferential analyses predata and postdata. *Psychol. Methods*. doi:10.1037/met0000697.

Vul, E., C. Harris, P. Winkielman, and H. Pashler. 2009. Puzzlingly high correlations in fMRI studies of emotion, personality, and social cognition. *Perspect. Psychol. Sci.* 4:274–290.

Wasserstein, R.L., and N.A. Lazar. 2016. The ASA Statement on p-Values: Context, Process, and Purpose. *Am. Stat.* 70:129–133.

Welch, B.L. 1938. The Significance of the Difference Between Two Means when the Population Variances are Unequal. *Biometrika*. 29:350–362.

Yang, Y., A. Sánchez-Tójar, R.E. O'Dea, D.W.A. Noble, J. Koricheva, M.D. Jennions, T.H. Parker, M. Lagisz, and S. Nakagawa. 2023. Publication bias impacts on effect size, statistical power, and magnitude (Type M) and sign (Type S) errors in ecology and evolutionary biology. *BMC Biol.* 21:71.